\RequirePackage{fix-cm}
\documentclass[smallextended]{svjour3}       
\smartqed  
\usepackage{cite}
\usepackage{amsfonts}
\usepackage{amsmath}
\usepackage{bbm}
\usepackage{float}
\usepackage[caption = false]{subfig}
\usepackage{graphicx}
\usepackage[round]{natbib}
\begin{document}

\title{Comparisons of Algorithms in Big Data Processing
}

\author{Amirali Daghighi         \and
        Jim Q. Chen 
}

\institute{Amirali Daghighi: adaghighi@stcloudstate.edu \at
              St. Cloud State University
           \and
           Jim Q. Chen: jchen@stcloudstate.edu \at
              St. Cloud State University
}
\date{}

\maketitle

\begin{abstract}
Change management of information systems includes careful assessment of 
increasing number of algorithms’ robustness. MapeReduce is a popular parallel computing 
framework for big data processing. 
Algorithms used in the framework prove to be effective 
only when certain conditions are true. The First-In-First-Out (FIFO) and
Hadoop Fair Scheduler (HFS) algorithms
do not take the rack structure of data centers into account, so they are
 known to not be heavy-traffic delay optimal or even throughput optimal. The recent advances on scheduling for data centers considering the rack 
structure and the heterogeneity of servers resulted in the state-of-the-art Balanced-PANDAS algorithm that outperforms the classic MaxWeight algorithm and its derivation,
 JSQ-MaxWeight algorithm. In both Balanced-PANDAS and MaxWeight algorithms, the processing rate of local, rack-local, and remote servers are assumed to be known. However, 
with the change of traffic over time in addition to estimation errors of processing rates, it is not realistic to consider the processing rates to be known.
 In this research, we study the robustness of Balanced-PANDAS and MaxWeight algorithms in
terms of inaccurate estimations of processing rates.
We observed that Balanced-PANDAS is not as sensitive as MaxWeight on the accuracy of processing rates, 
making it more appealing to use in Big Data processing centers.

\end{abstract}

\section{Introduction and Related Work}
\label{intro}

Parallel computing for big data has different applications from online social networks, health-care industry, advertisement placement, and machine learning to search engines and data centers. The most popular big data parallel computing framework is MapReduce which is broadly used in Hadoop \citep{white2012hadoop}, Dryad \citep{isard2007dryad}, Google \citep{dean2008mapreduce}, Deep Learning \citep{daghighi2019application}, and grid-computing \citep{isard2009quincy, saadatmand2019dual, saadatmand2019heuristic}.
Before talking about MapReduce, we present some details on the network structure of data centers. Data centers used to mainly consist of two parts, storage and computing, and these two parts were connected to each other through a network link with high bandwidth. When this structure was being used for normal data processing, the communication of data from storage to processing unit was not creating any bottleneck. However, with the emergence of Big Data, the network between the two units was unable to accommodate fast and reliable data transfer. Hence, scientists come up with the idea of bringing data into processing unit by splitting both data and processing units into hundreds of small units and combining each computing unit with a small storage unit. As a result, each small unit consisting of storage and processing units, called a server, is capable of storing small pieces of data and processing them at the same time. In other words, data does not need to be transferred from storage unit to processing unit since they are already together. However, note that the big data cannot be stored in the storage unit of a single server. The solution is to split the big data into small chunks of data, typically 68-128 MB, and storing them on multiple servers. In practice, each data chunk is stored on three servers to increase availability and decrease data loss probability. As a result, processing of big data consists of processing of multiple data chunks that make the whole big data, and concatenating the data chunk processing results for the completion of the big data process. The processing of each data chunk is called Map task and the concatenation of the results on all the Map tasks is called Reduce task, which make up the MapReduce processing framework for big data. Note that at least for Reduce tasks, servers need to be connected and cannot completely be isolated. We later see that even for executing Map tasks, servers may need to exchange data. Hence, servers are connected to each other through links and switches. The structure of switches connecting servers is a complete field of research in computer science, but servers are generally connected to each other by top of the rack switches as well as core switches in the following way. The hundred servers are grouped into batches of 20-50 servers, where each batch is inter-connected to each other by a switch, called rack switch, and all rack switches are connected to one or more core switches which make all the servers connected.

The rack structure of data centers brings a lot of complexity for load balancing. As a result, most theoretical work on load balancing for data centers either consider homogeneous model of servers or ignore the rack structure and only consider data locality, where data locality refers to the fact that the data chunk associated to a Map task is stored on three servers, so is not available on other servers immediately. Examples of works that consider homogeneous server model are \citep{livny1982load, singh1970two, kreimer2002real}, and \citep{lowery2008systems}.
A branch of research on homogeneous servers is utilization of the power of two or more choices for load balancing that lowers the messaging overhead between hundreds of servers and the core load balancing scheduler.
For example you can refer to \citep{dahlgaard2016power, mitzenmacher2001power, gardner2017redundancy, richa2001power, byers2004geometric, doerr2011stabilizing, lumetta2007using, cooper2014power, luczak2005power}, and \citep{gast2015power}.
Although there has been a huge body of work on heuristic algorithms for heterogeneous server model, examples of which are \citep{apache, isard2009quincy, zaharia2010delay, jin2011bar, he2011matchmaking, polo2011resource, zaharia2008improving, daghighi2017scheduling, kavousi2017affinity, moaddeli2019power}, and \citep{salehi2017optimal}, there has been a few recent works on algorithms with theoretical guarantees on such more complicated models.

The scheduling problem for a data center with a rack structure is a specific case of the open affinity scheduling problem, where each task type can be processed by each server but with different processing rates.
The classic MaxWeight algorithm \citep{van2009instability, sadiq2009throughput, stolyar2004maxweight, meyn2009stability}, c-$\mu$-rule \citep{van1995dynamic, mandelbaum2004scheduling}, and the work by \cite{harrison1999heavy, harrison1998heavy}, and by \cite{bell2001dynamic, bell2005dynamic} have different approaches on solving the load balancing problem for the affinity scheduling, but they either not solve the delay optimality or have unrealistic assumptions including known task arrival rates and existence of one queue per task type.
The state-of-the-art on scheduling for data centers considering the rack structure, no knowledge of task arrival rates, and having queues on the order of servers not the number of task types is presented by \cite{xie2016scheduling} and by \cite{yekkehkhany2017near}, which is extended for a general number of data locality levels by \cite{yekkehkhany2018gb}.
The central idea to all algorithms in \citep{xie2016scheduling, yekkehkhany2017near}, and \citep{yekkehkhany2018gb} is to use weighted workload on servers instead of the queue lengths, which leads to a better perfromance in terms of average delay expereinced by submitted tasks. The Balanced-PANDAS alborithm, where PANDAS stands for Priority Algorithm for Near-Data Scheduling, is the name for the weighetd-workload based algorithms proposed in \citep{xie2016scheduling, yekkehkhany2017near}, and \citep{yekkehkhany2018gb}.
The Join-the-Shortest-Queue-MaxWeight (JSQ-MaxWeight, JSQ-MW) proposed by  \cite{wang2013throughput} that only considers data locality is also extended to the case where rack structure is considered by \cite{xie2016scheduling}. The priority algorithm proposed by \cite{xie2015priority} is another work that only considers data locality, not the rack structure of data centers which is interesting in its own rights since both throughput and heavy-traffic delay optimality are proved for it; however, it is not even throughput optimal for a system with rack structure.

All of the algorithms mentioned above consider complete knowledge about the processing rates of different task types on different servers. However, the reality is that the processing rates are mostly not known due to errors in estimation methods and the change of the system structure over time or the change of traffic which can change the processing rates. Hence, it is important that the algorithm that is used for load balancing is robust to estimation errors of processing rates. In this work, we run extensive simulations to evaluate the robustness of the state-of-the-art algorithms on load balancing with different levels of data locality. It is observed that the Balanced-PANDAS algorithm not only has a better heavy-traffic delay performance, but it also is more robust to changes of processing rate estimations, while MaxWeight based algorithm does not perform as well as Balanced-PANDAS under processing rate estimation errors.
In order to estimate the processing rates of tasks on servers and better model the data center structure, reinforcement learning methods can be used as it is discussed in  \citep{musavi2016game, yildiz2013predicting, musavi2016unmanned}, and \citep{musavi2019game}.
A recent work by \cite{yekkehkhany2020blind} considers an exploration-exploitation approach as in the reinforcement learning method to both learn the processing rates and exploit load balancing based on the current estimation of the processing rates. They propose the Blind GB-PANDAS algorithm that is proven to be throughput-optimal and have a lower mean task completion time than the existing methods. A more sophisticated risk-averse exploration-exploitation approach can be considered for this problem when different tasks have different risk-levels as discussed in \citep{yekkehkhany2019risk} and \citep{yekkehkhany2020cost}.

The rest of the paper is organized as follows. Section \ref{sec:1} presents the system model that is used throughout the paper, section \ref{sec:2} summarizes the preliminary materials including the description of Balanced-PANDAS, Priority, and MaxWeight based algorithms that are needed before we present the robustness comparison among different algorithms in section \ref{sec:3}.

\section{System Model}
\label{sec:1}
We consider the same system model described in \citep{xie2016scheduling} and \citep{yekkehkhany2018gb} as follows. A discrete time model is considered, where time is indexed by $t \in \mathbb{N}$. Assume a data center with $M$ servers and denote the set of all servers as $\mathcal{M} = \{1, 2, \cdots, M\}$. Without loss of generality, assume that the first $M_R$ servers are connected to each other with a top of the rack switch and are called the first rack, the second $M_R$ servers are connected to each other with another top of the rack switch and are called the second rack, and so on. Hence, there are $N_R = \frac{M}{M_R}$ racks in total. All the top of the rack switches are connected to each other with one or more core switches in a symmetric manner. As a result, there are three levels of data locality as described below. Recall that the data chunk associated to a map task is stored on three servers by Hadoop's default, so all those three servers are called local servers for the map task or in other words the map task can receive service locally from those three servers. Since servers have the data chunk of local tasks, the processing is immediately started after servers are assigned to process them. Note that the three servers storing the data for a map task is normally different for different map tasks. Hence, we associate a type to each map task, which is the label of the three servers, i.e. $(m_1, m_2, m_3) \in \mathcal{M}^3,$ such that $m_1 < m_2 < m_3$. This gives us a unique and informative way of representation for different task types as follows:
$$\bar{L} \in \mathcal{L} = \{ (m_1, m_2, m_3) \in \mathcal{M}^3: m_1 < m_2 < m_3 \},$$
where a task type is denoted by $\bar{L} = (m_1, m_2, m_3)$ given that $m_1, m_2,$ and $m_3$ are the three local servers for task of type $\bar{L}$ and the set of all task types is denoted by $\mathcal{L}$.

A map task is not limited to receive service from one of the local servers. It can receive service from one of the servers that are in the same rack as the local servers with a slightly lower service rate. The slight depreciation of processing rate for such servers, which are called rack-local servers, is for the travel time of the data associated to a map task from a local server to the rack-local server that is assigned for processing the map task rack-locally. Finally, all other servers other than the local and rack-local servers, which are called remote servers, have the lowest processing rate for a map task, since data needs to be transmitted through at least two of the top rack switches and a core switch, so the server cannot immediately start processing the task when it is assigned to do so remotely. In order to formally define the rack-local and remote servers, we need to propose a notation. Let $R(m) \in \{1, 2, \cdots, N_R\}$ denotes the label of the rack that the $m$-th server belongs to. Then, the set of rack-local and remote servers to task of type $\bar{L} = (m_1, m_2, m_3)$, denoted by $\bar{L}_k$ and $\bar{L}_r$, respectively, are as follows:
$$\bar{L}_k = \Big \{m \in \mathcal{M} : m \not \in (m_1, m_2, m_3), R(m) \in \Big (R(m_1), R(m_2), R(m_3) \Big ) \Big \},$$
$$\bar{L}_r = \Big \{m \in \mathcal{M} : R(m) \not \in \Big (R(m_1), R(m_2), R(m_3) \Big ) \Big \}.$$

The service and arrival process of tasks is described below. \\
\textbf{Task arrival process:} Let $A_{\bar{L}}(t)$ denote the number of tasks of type $\bar{L}$ that arrive to the system at time slot $t$, where $\mathbb{E}[A_{\bar{L}}(t)] = \lambda_{\bar{L}},$ and it is assumed that $A_{\bar{L}}(t) < C_A$ and $P(A_{\bar{L}}(t) = 0) > 0$. The set of arrival rates for all task types is denoted by $\mathbf{\lambda} = (\lambda_{\bar{L}}: \bar{L} \in \mathcal{L})$. \\
\textbf{Service process:}
The processing of a task on a local server is assumed to be faster than on a rack-local server, and the processing of a task on a rack-local server is faster than on a remote server. This fact is formalized as follows. The processing time of a task on a local, rack-local, and remote server has means $\frac{1}{\alpha}, \frac{1}{\beta},$ and $\frac{1}{\gamma},$ respectively, where $\alpha > \beta > \gamma.$ Note that the processing time of a task can have any distribution with the given means, but the heavy-traffic delay optimality of Balanced-PANDAS algorithm is only proven under Geometric service time distribution, while MaxWeight based algorithm does not have a general heavy-traffic delay optimality under any distribution for service time. \\
\textbf{Capacity region characterization:}
An arrival rate for all task types is supportable for service by the $M$ servers if and only if the load on each server is strictly less than the capacity of the server. Considering a processing rate of one for each server, an arrival rate vector $\mathbf{\lambda} = (\lambda_{\bar{L}}: \bar{L} \in \mathcal{L})$ is in the capacity region of the system if and only if:
$$\sum_{\bar{L}: m \in \bar{L}} \frac{\lambda_{\bar{L}, m}}{\alpha} + \sum_{\bar{L}: m \in \bar{L}_k} \frac{\lambda_{\bar{L}, m}}{\beta} + \sum_{\bar{L}: m \in \bar{L}_r} \frac{\lambda_{\bar{L}, m}}{\gamma} < 1, \ \forall m \in \mathcal{M},$$
where $\lambda_{\bar{L}, m}$ is the rate of incoming tasks of type $\bar{L}$ that are processed by server $m$.

\section{Load Balancing Algorithms}
\label{sec:2}
In this section, we introduce the main three algorithms on scheduling for data centers with more than or equal to two levels of data locality that have theoretical guarantees on optimality in some senses and under some conditions. In order to introduce the load balancing algorithm of each method, we also need to present the queueing structure required for that method. The three algorithms are
\begin{enumerate}
  \item Priority algorithm \citep{xie2015priority}, which is best fit for applications with two levels of data locality, e.g. for the cases that only data locality is taken into account. An example is scheduling for Amazon Web Services inside a rack.
  \item Balanced-PANDAS \citep{xie2016scheduling, yekkehkhany2017near}, and \citep{yekkehkhany2018gb}, which is the state-of-the-art for scheduling applications with multiple levels of data locality and is observed to perform better in terms of average task completion time by fourfold in comparison to MaxWeight based algorithms. It is proven by \citep{xie2016scheduling} that under mild conditions, Balanced-PANDAS is both throughput and heavy-traffic delay optimal for a system with three levels of data locality and a rack structure.
  \item MaxWeight based algorithms \citep{stolyar2004maxweight} and \citep{wang2016maptask}, which can be used for multiple levels of data locality and are throughput optimal, but not heavy-traffic delay optimal, and it is observed that they generally have poor performance at high loads compared to weighted workload based algorithm used in Balanced-PANDAS algorithm.
\end{enumerate}
The following three subsections present a complete introduction to these three main algorithms.

\subsection{Priority algorithm}
The Priority algorithm is designed for a system with two levels of data locality. In other words, it only considers data locality, but not the rack structure. Hence, there are only local and remote servers from the perspective of a task. The queueing structure under this algorithm is to have a single queue per server, where the queue corresponding to a server only keeps tasks that are local to that server. At the arrival of a task, a central scheduler routes the incoming task to the local server with the shortest queue length. An idle server is scheduled to process a task in its corresponding queue as long as there is one, and if the idle server's queue length is zero, it is scheduled to process a task from the longest queue in the system. The priority algorithm is proved to be both throughput and heavy-traffic delay optimal. However, its extension to more than two levels of data locality is not even throughput optimal, let alone heavy-traffic delay optimal.
\subsection{Balanced-PANDAS algorithm}
The Balanced-PANDAS algorithm can be used for a system with multiple levels of data locality, but here we propose the algorithm for a data center with a rack structure with three levels of data locality. The queueing structure under using this algorithm is to have three queues per server, one queue for storing local tasks to the server, another queue for storing rack-local tasks to the server, and a third queue for storing remote tasks to the server. Hence, server $m$ has a tuple of three queues denoted by $\left (Q_m^l, Q_m^k, Q_m^r \right )$, where they refer to the queues storing local, rack-local, and remote tasks respectively. The corresponding queue lengths at time $t$ are denoted by $\left (Q_m^l(t), Q_m^k(t), Q_m^r(t) \right )$. The workload on server $m$ at time slot $t$ is defined as follows:
$$W_m(t) = \frac{Q_m^l(t)}{\alpha} + \frac{Q_m^k(t)}{\beta} + \frac{Q_m^r(t)}{\gamma}.$$
An incoming task of type $\bar{L}$ is routed to the corresponding queue of the server with the minimum weighted workload, where ties are broken randomly, in the set below:
$$\underset{m \in \mathcal{M}}{\arg\min} \left \{ \frac{W_m(t)}{\alpha \cdot \mathbbm{1}_{\{ m \in \bar{L} \}} + \beta \cdot \mathbbm{1}_{\{ m \in \bar{L}_k \}} + \gamma \cdot \mathbbm{1}_{\{ m \in \bar{L}_r \}}} \right \}.$$
An idle server $m$ at time slot $t$ is scheduled to process a local task from $Q_m^l$ if $Q_m^l(t) \neq 0$; otherwise, it is scheduled to process a rack-local task from $Q_m^k$ if $Q_m^k(t) \neq 0$; otherwise, it is scheduled to process a remote task from $Q_m^r$ if $Q_m^r(t) \neq 0$; otherwise, it remains idle until a task joins one of its three queues. The Balanced-PANDAS algorithm is throughput optimal. It is also heavy-traffic delay optimal for a system with a rack structure of three levels of data locality if $\beta^2 > \alpha \cdot \gamma$, which means that the rack-local service is faster than the remote service in a specific manner. 
\subsection{MaxWeight based algorithm}
The MaxWeight algorithm is proposed by \cite{stolyar2004maxweight} and a modification of it called JSQ-MaxWeight algorithm is proposed by \cite{wang2016maptask}, which is described below.
Consider one queue per server, i.e. server $m$ has a single queue called $Q_m$, where its queue length at time slot $t$ is denoted by $Q_m(t)$. The routing policy is as the Priority algorithm, i.e. an incoming task of type $\bar{L}$ is routed to the queue of the shortest length in $\bar{L}$. This routing policy is called join-the-shortest-queue (JSQ). An idle server $m$ at time slot $t$ on the other hand is scheduled to process a task from a queue with the maximum weighted queue length, where ties are broken at random, in the set below:
$$\underset{n \in \mathcal{M}}{\arg \max} \left \{ \left ( \alpha \cdot \mathbbm{1}_{\{n = m\}} + \beta \cdot \mathbbm{1}_{\{n \neq m, R(n) = R(m)\}} + \gamma \cdot \mathbbm{1}_{\{R(n) \neq R(m)\}} \right ) \cdot Q_n(t) \right \}.$$
The JSQ-MaxWeight algorithm is throughput optimal, but it is not heavy-traffic delay optimal.

\section{Robustness Comparison of Scheduling Algorithms}
\label{sec:3}
\begin{figure*}
\center
  \includegraphics[width=0.75\textwidth]{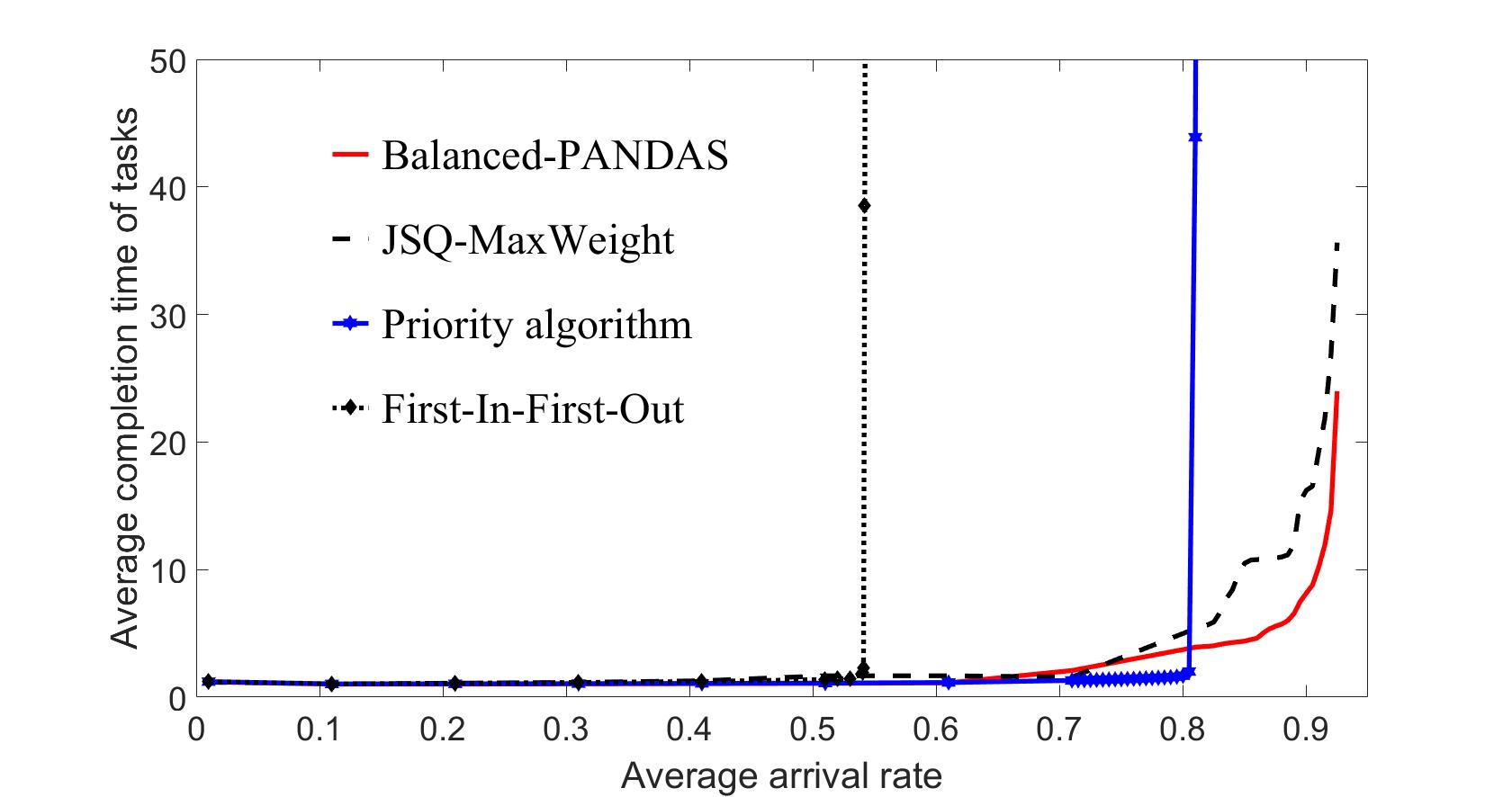}
\caption{Comparison of the algorithms using the precise value of parameters.}
\label{fig:1}
\end{figure*}

\begin{figure*}
\center
  \includegraphics[width=0.75\textwidth]{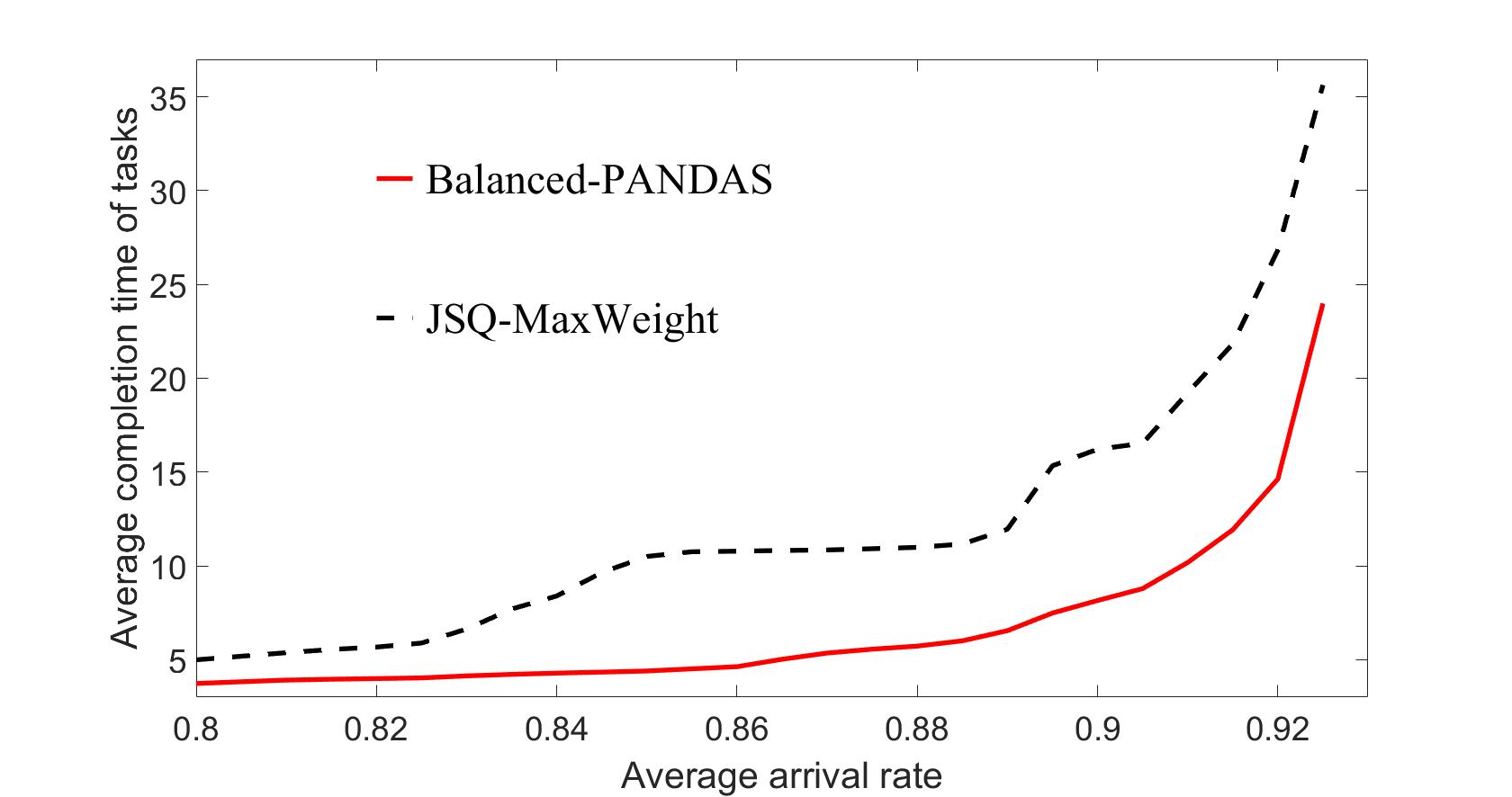}
\caption{Comparison of Balanced-PANDAS and JSQ-MaxWeight at high loads using the precise value of parameters.}
\label{fig:2}
\end{figure*}

In this section, we present the results on our extensive simulations on robustness of scheduling algorithms presented in section \ref{sec:2}. To this end, we run the algorithms with parameters that have error to study which algorithm can tolerate errors better than others. More specifically, we use incorrect $\alpha, \beta,$ and $\gamma$ in the algorithms for calculating weighted workloads or weighted queue lengths and observe the average task completion time under these scenarios. The arrival process is a Poisson process and the processing time has an exponential distribution. We have also tested the algorithms for processing times with heavy-tailed distributions and observed similar results. We make these parameters $5\%, 10\%, 15\%, 20\%, 25\%,$ and $30\%$ off their real value, either greater than the real value or smaller than the real value, and evaluate algorithms under these cases. The traffic load is assumed the same under all algorithms so that the comparison makes sense. We further compare all the three algorithms mentioned in section \ref{sec:2} with the Hadoop's default scheduler which is First-In-First-Out (FIFO). Figure \ref{fig:1} shows the comparison between the four algorithms when precise value of parameters are known by the central scheduler. As we see, Balanced-PANDAS algorithm has the lowest average task completion time at high loads. A closer look of high loads is presented in figure \ref{fig:2}, where Balanced-PANDAS obviously outperform JSQ-MaxWeight in terms of average task completion time.

\begin{figure}
\center
\subfloat[Parameters are off for 5\% lower]{\includegraphics[width = 2.495in]{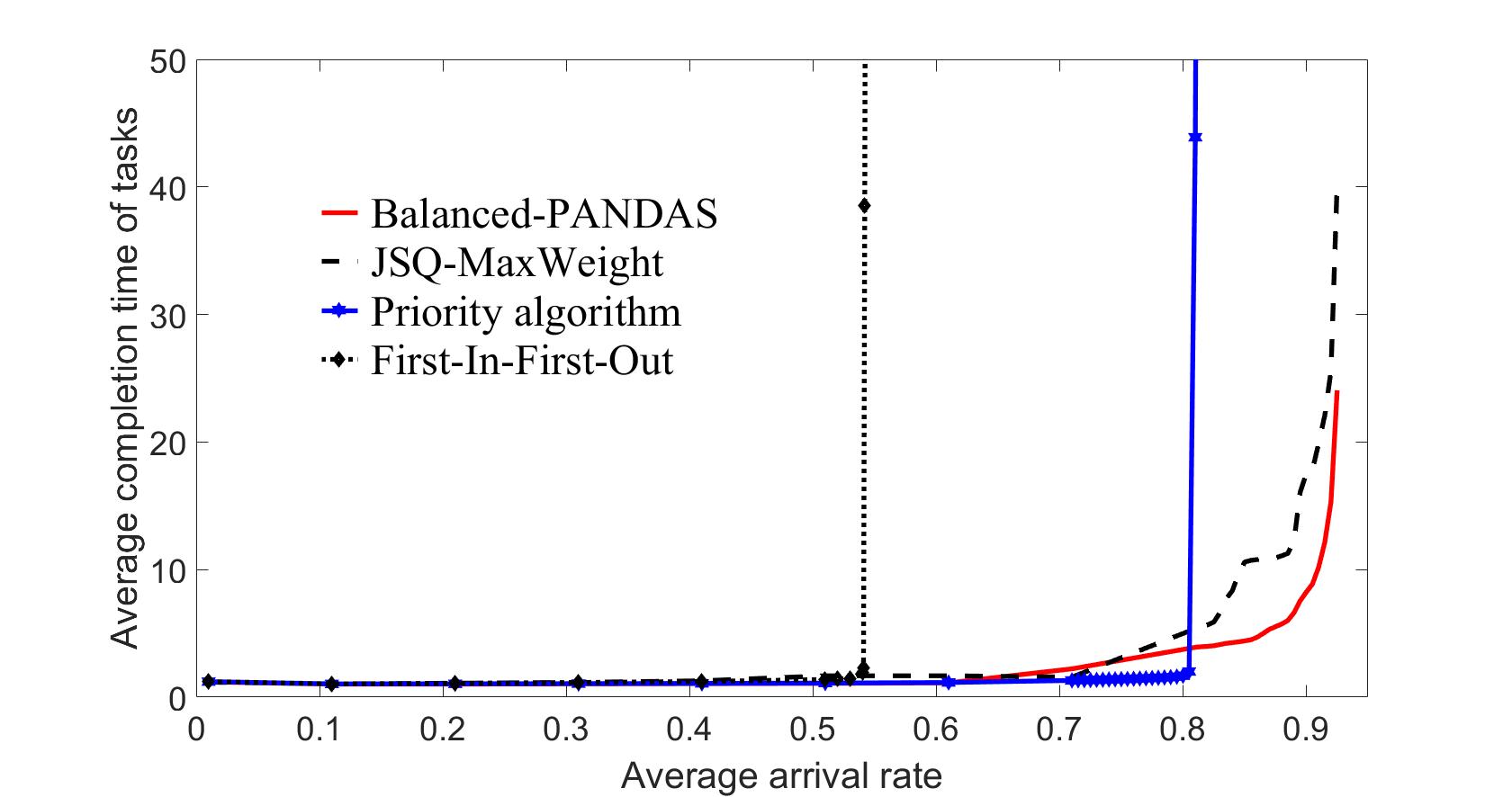}}
\subfloat[Parameters are off for 10\% lower]{\includegraphics[width = 2.495in]{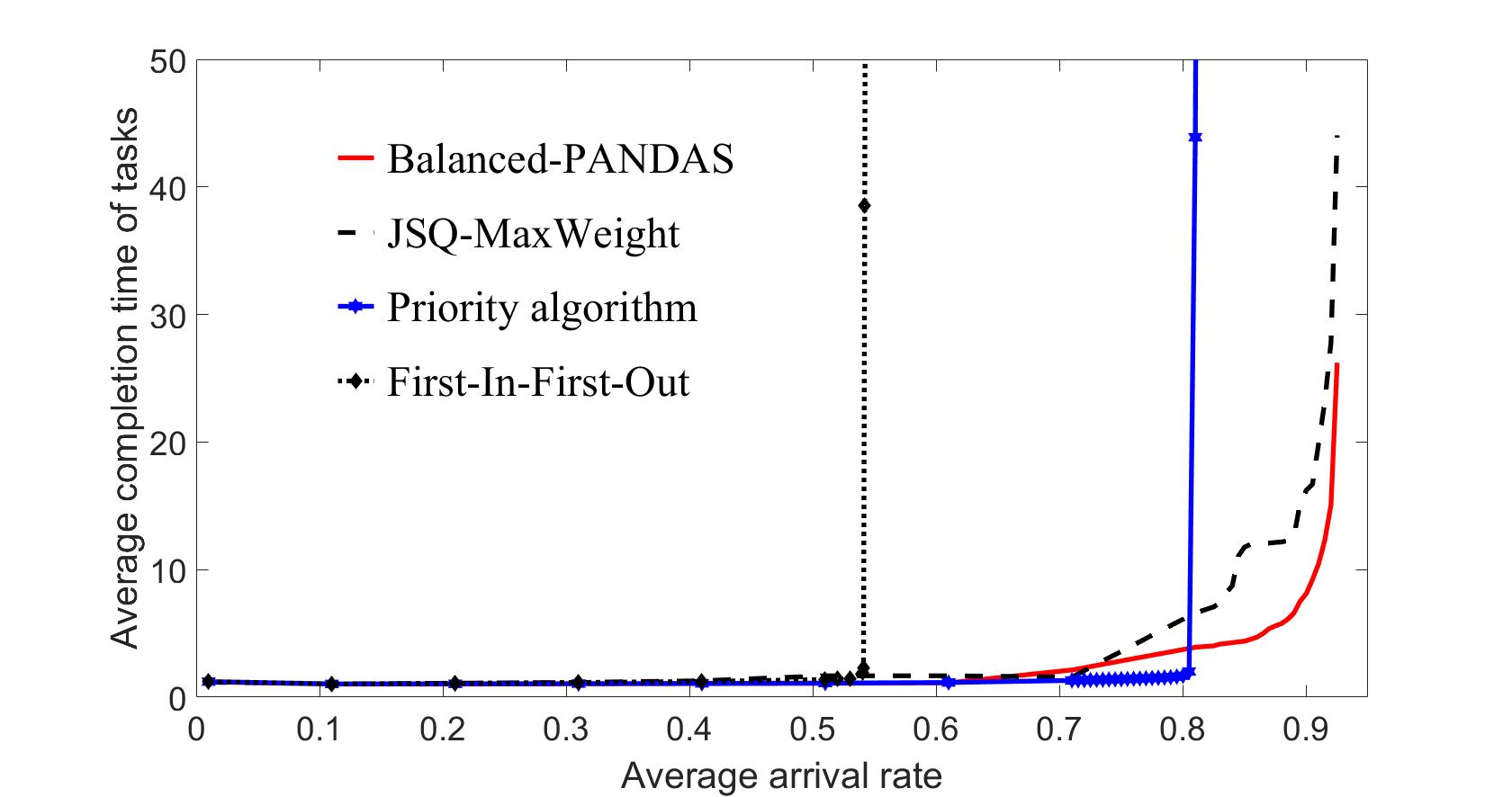}} \\
\subfloat[Parameters are off for 15\% lower]{\includegraphics[width = 2.495in]{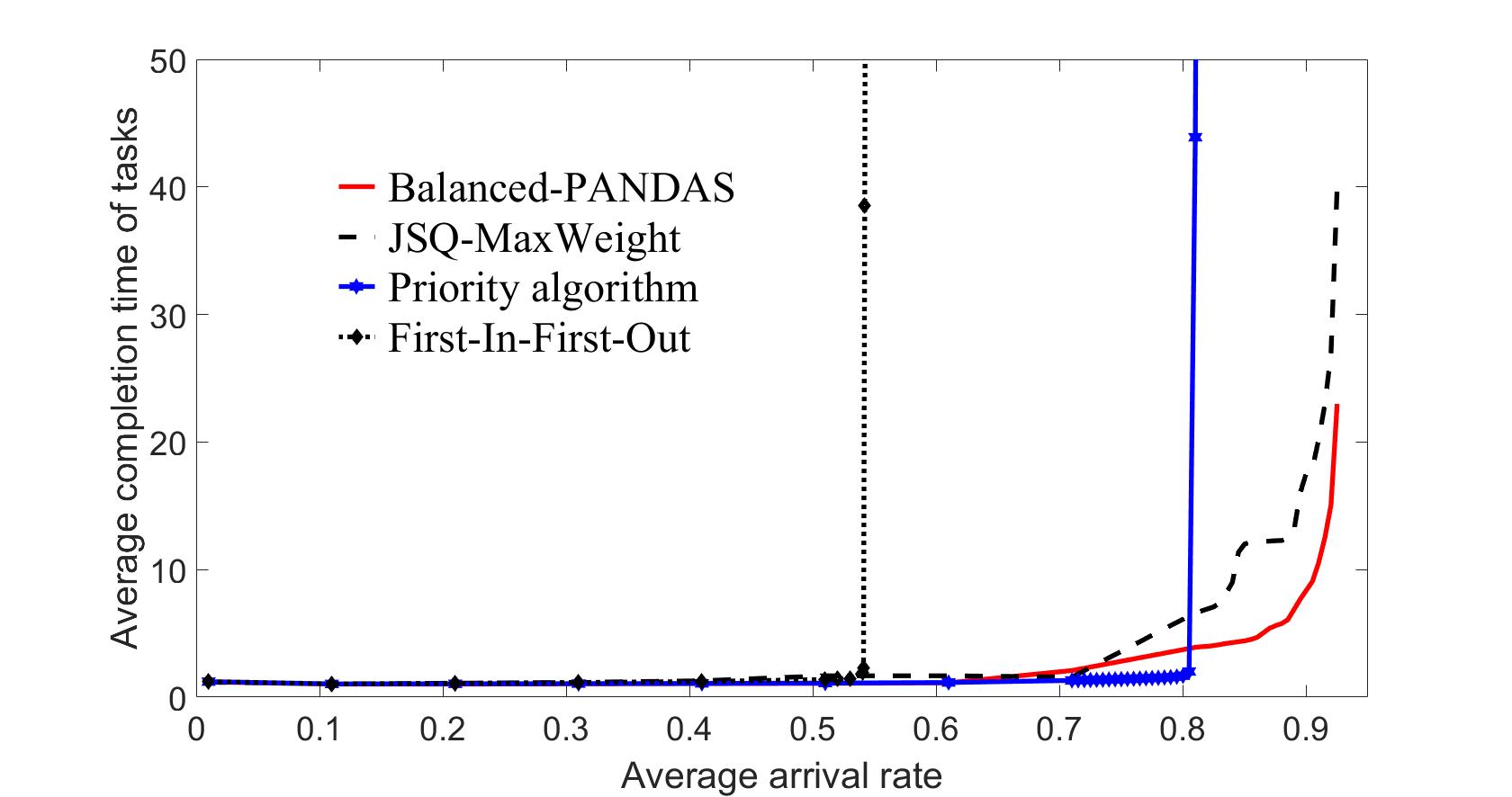}}
\subfloat[Parameters are off for 20\% lower]{\includegraphics[width = 2.495in]{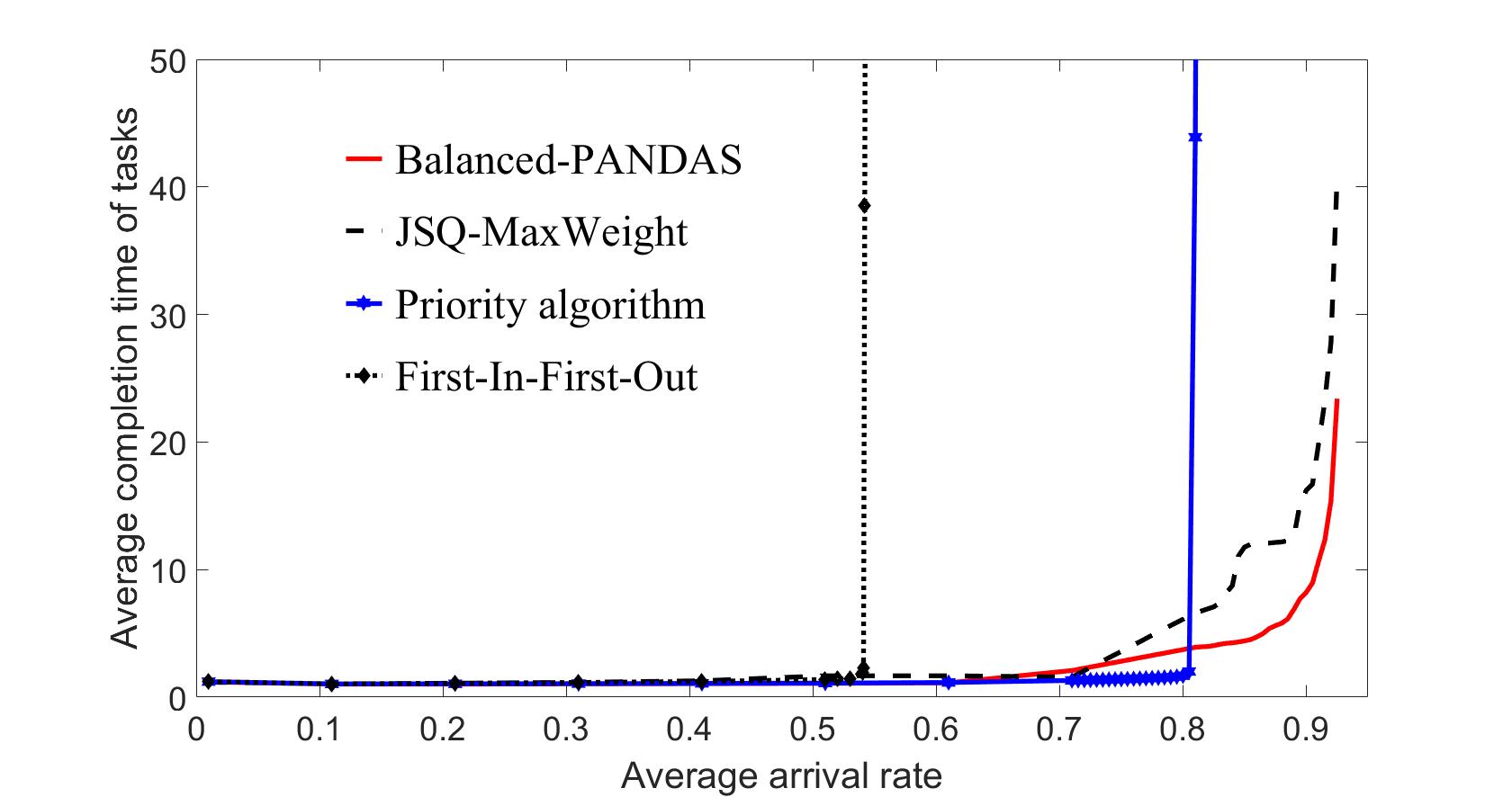}} \\
\subfloat[Parameters are off for 25\% lower]{\includegraphics[width = 2.495in]{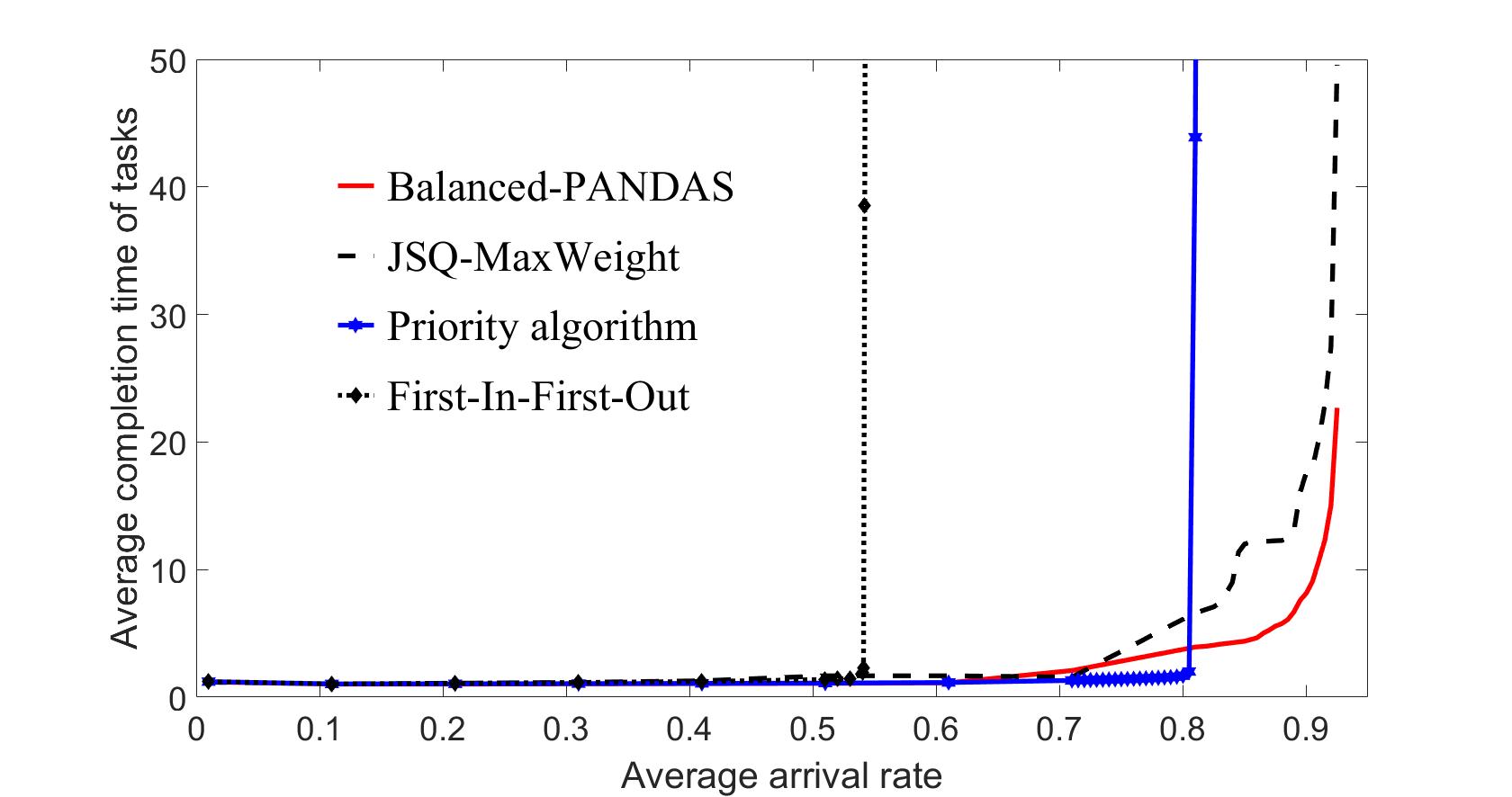}}
\subfloat[Parameters are off for 30\% lower]{\includegraphics[width = 2.495in]{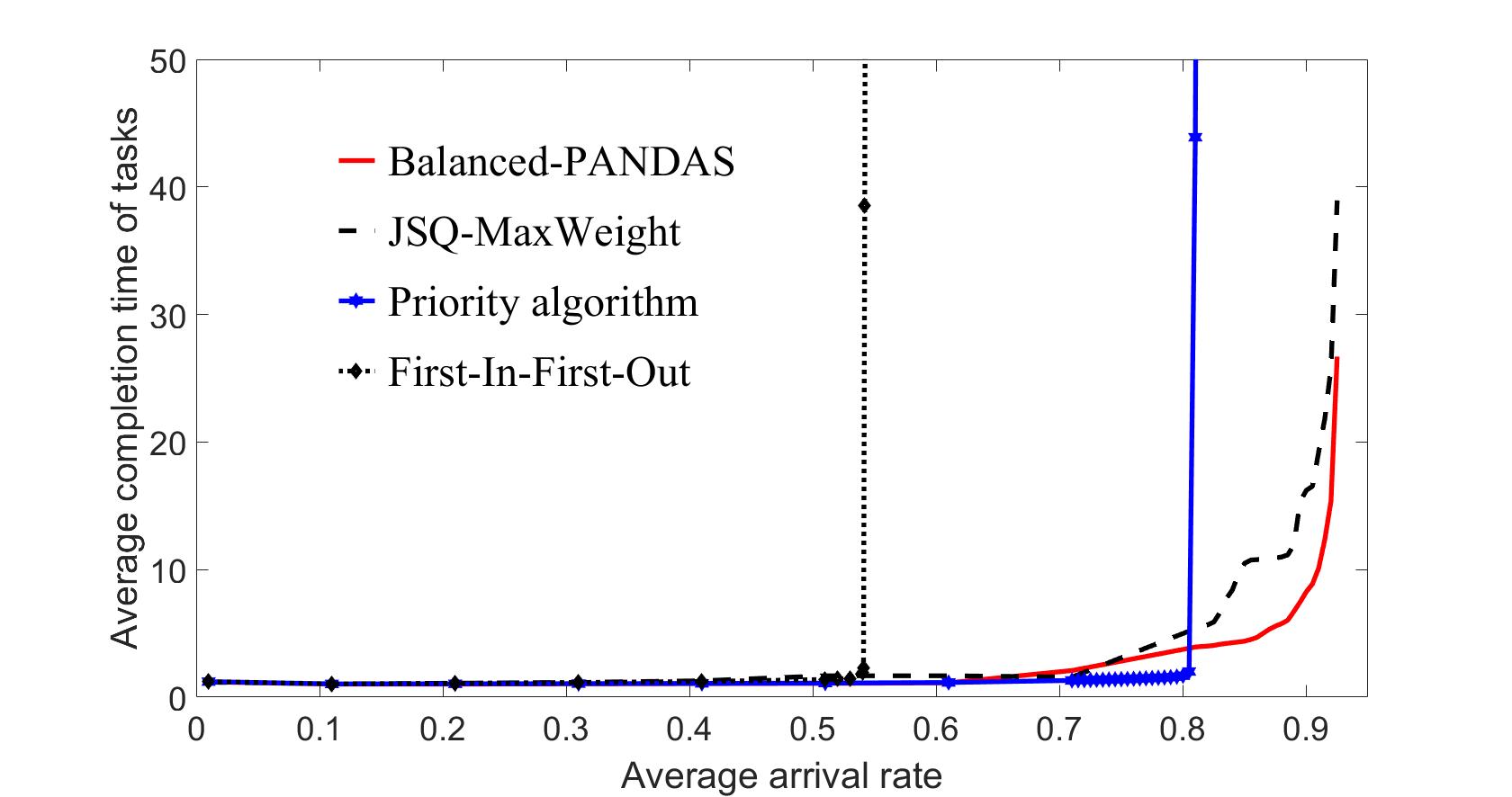}} 
\caption{Robustness comparison of algorithms when parameters are off and lower than their real values.}
\label{fig:3}
\end{figure}

\begin{figure*}
\center
  \includegraphics[width=0.75\textwidth]{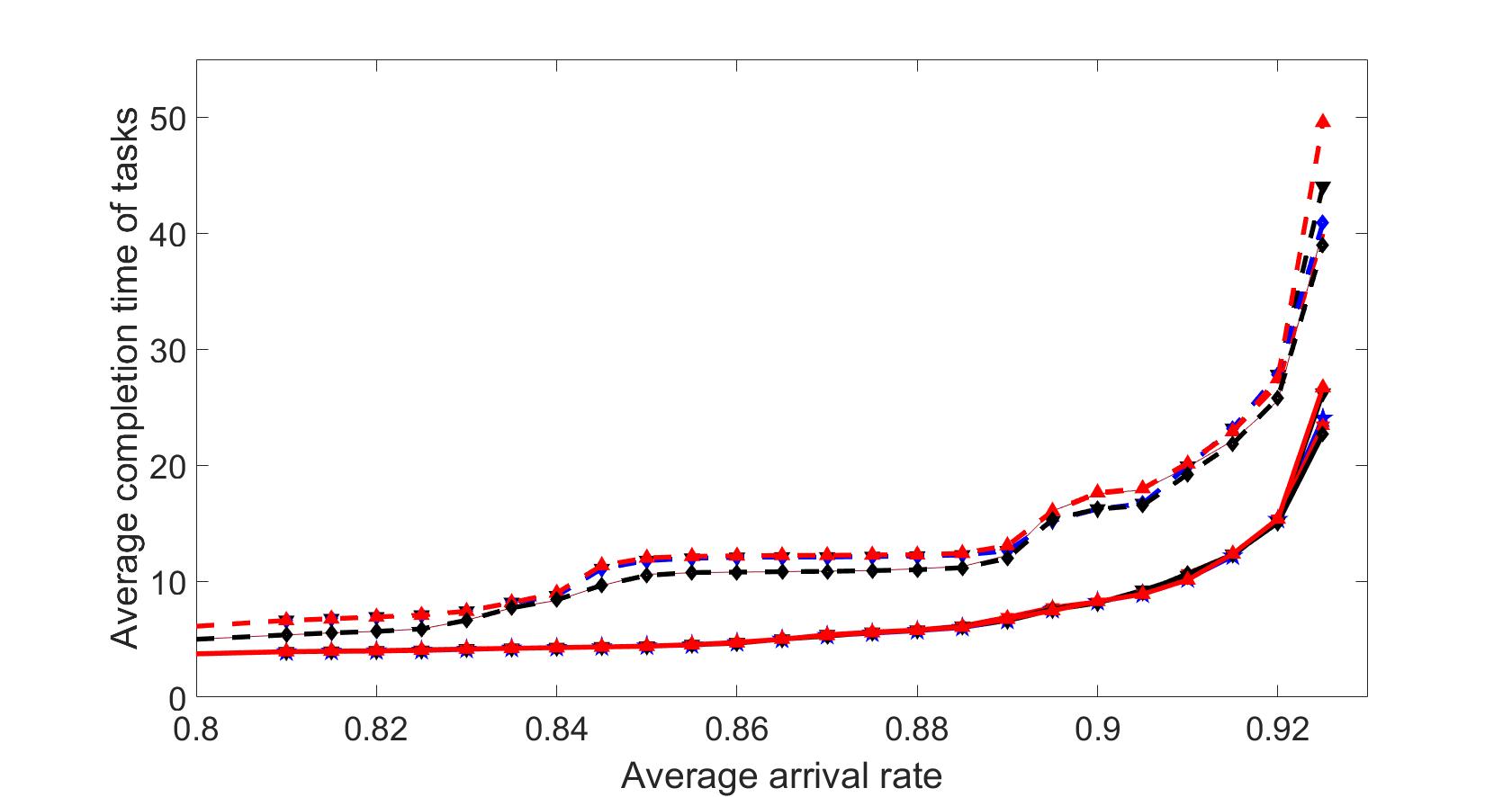}
\caption{Sensitivity comparison of Balanced-PANDAS and JSQ-MaxWeight against parameter estimation error.}
\label{fig:4}
\end{figure*}

The performance of the four algorithms are compared to each other when the parameters are lower than their real values by certain percentages, where the results are shown in figure \ref{fig:3}. As is seen, Balanced-PANDAS has best performance among all algorithms by changing the parameters' error from 5\% to 30\%. In fact, figure \ref{fig:4} shows that the Balanced-PANDAS has the least sensitivity against change of parameters while JSQ-MaxWeight's performance varies notably by the increase of error in parameter estimations.

\begin{figure}
\center
\subfloat[Parameters are off for 5\% higher]{\includegraphics[width = 2.495in]{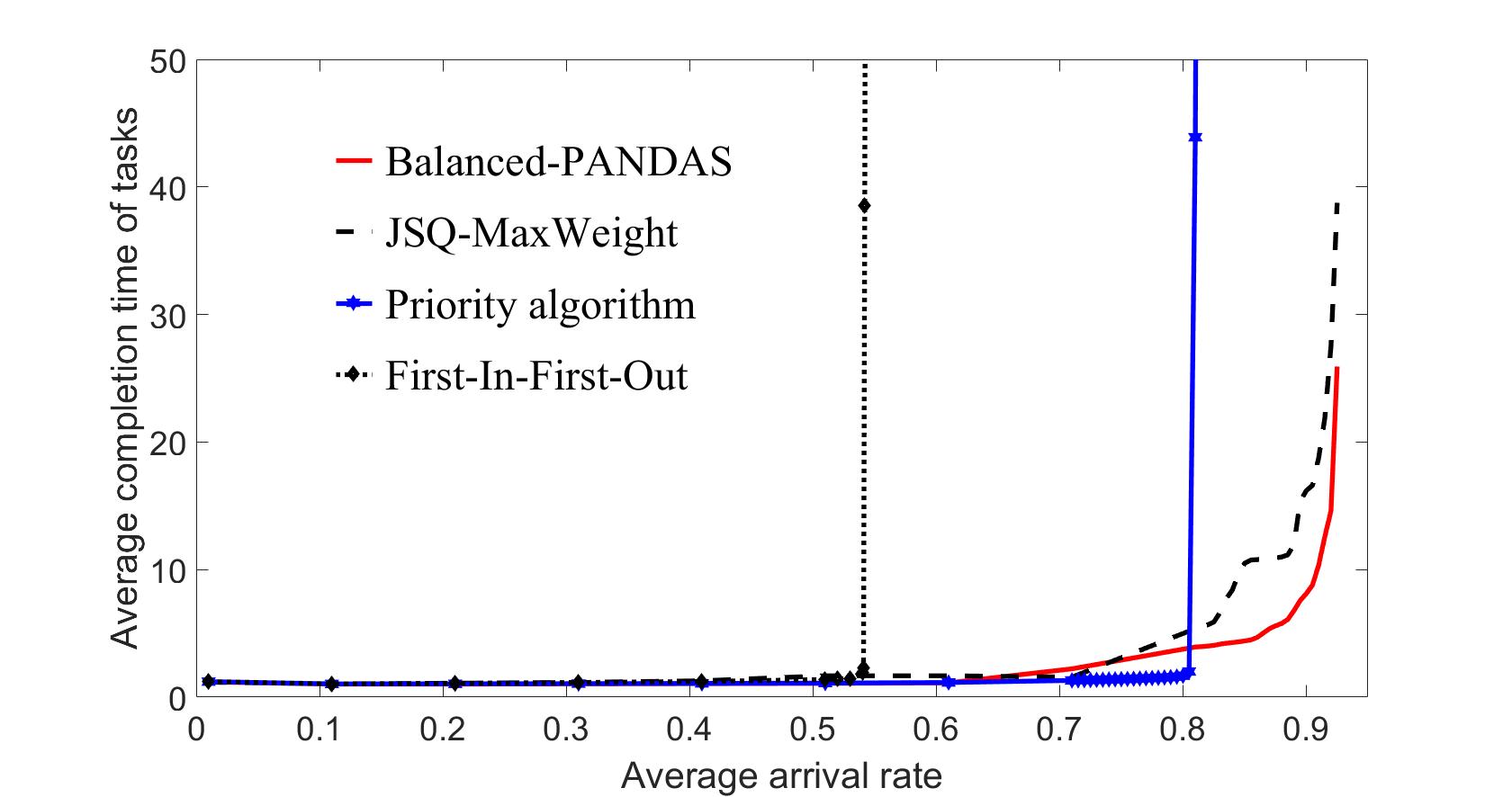}}
\subfloat[Parameters are off for 10\% higher]{\includegraphics[width = 2.495in]{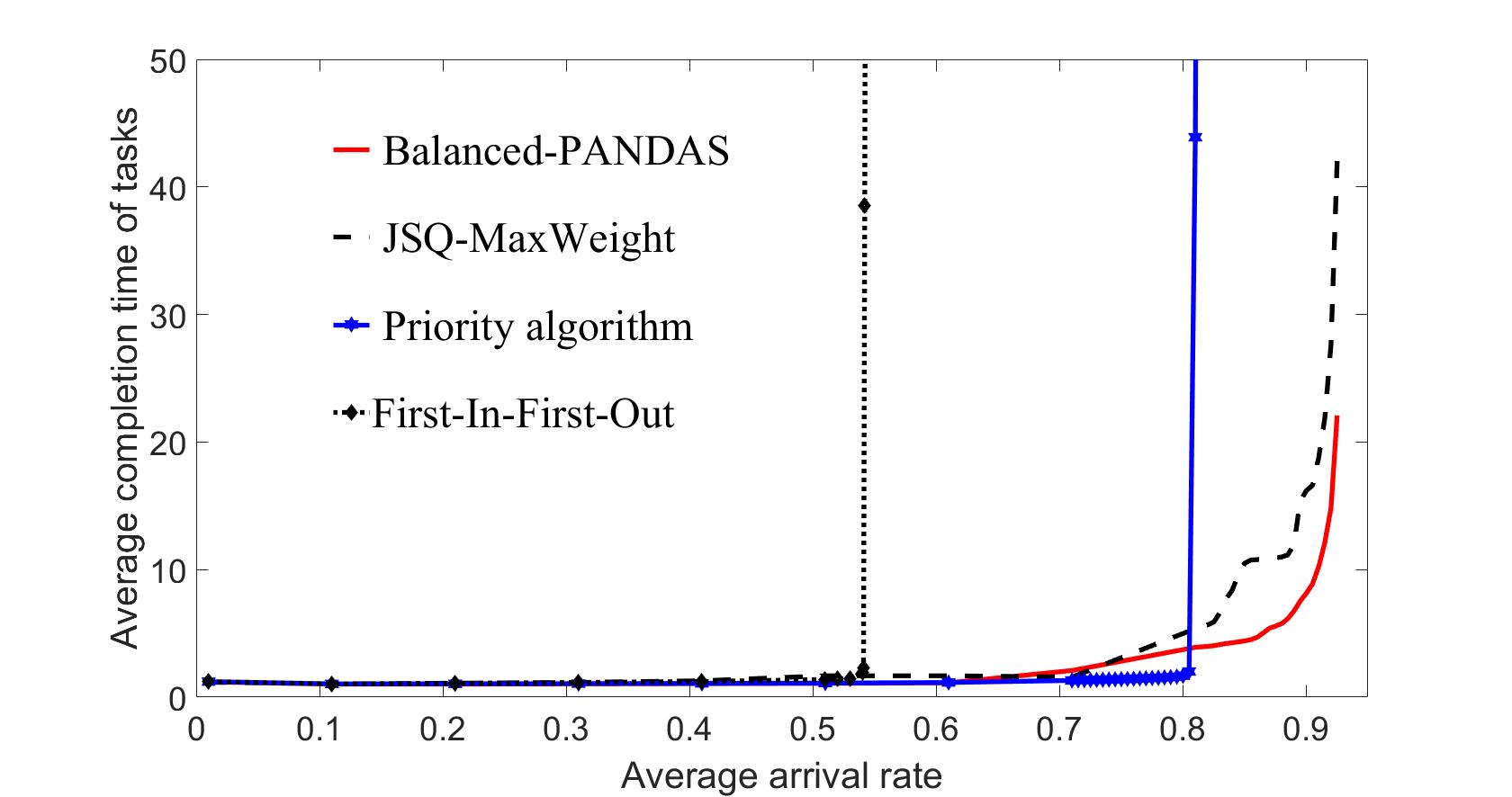}} \\
\subfloat[Parameters are off for 15\% higher]{\includegraphics[width = 2.495in]{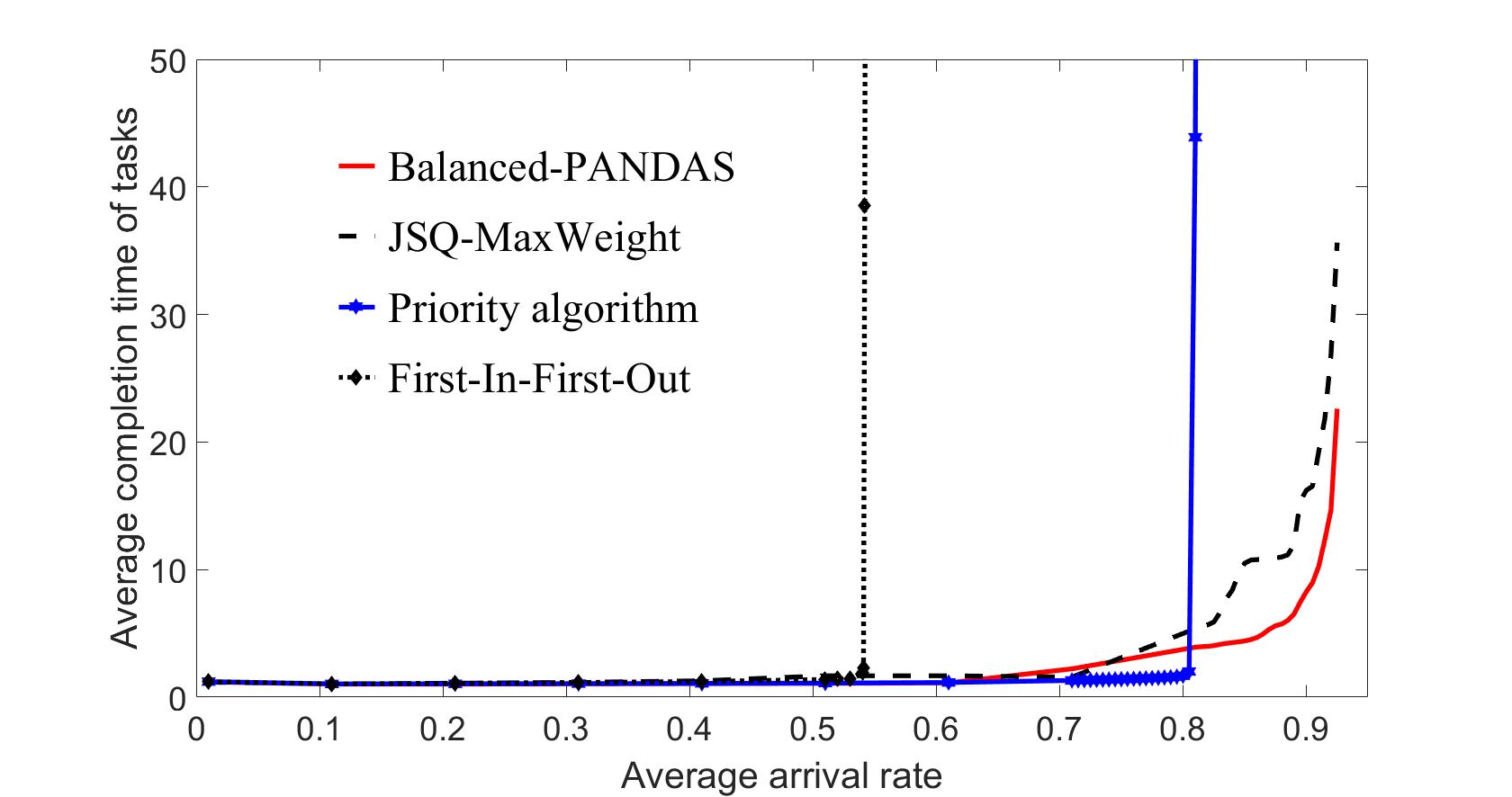}}
\subfloat[Parameters are off for 20\% higher]{\includegraphics[width = 2.495in]{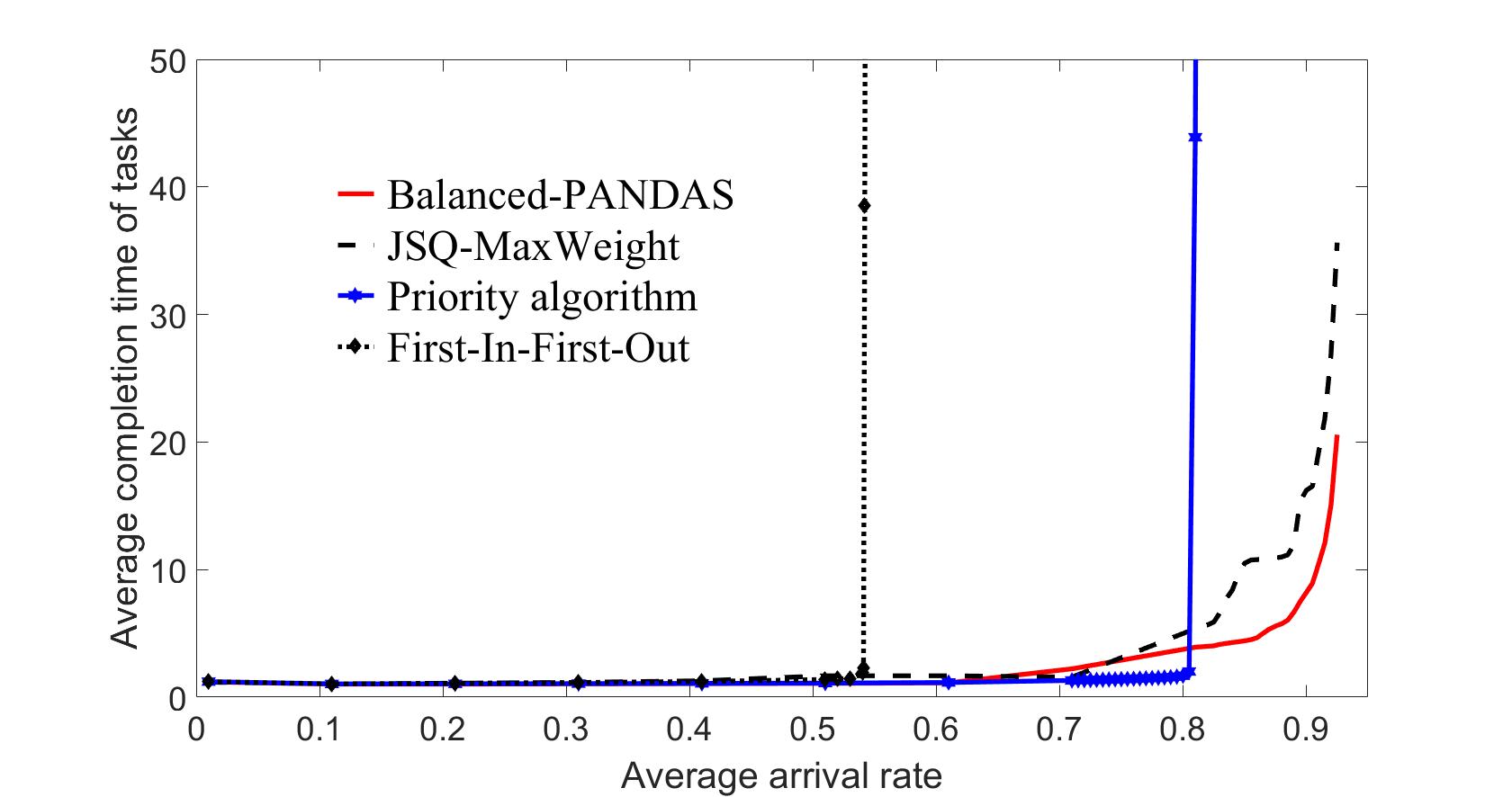}} \\
\subfloat[Parameters are off for 25\% higher]{\includegraphics[width = 2.495in]{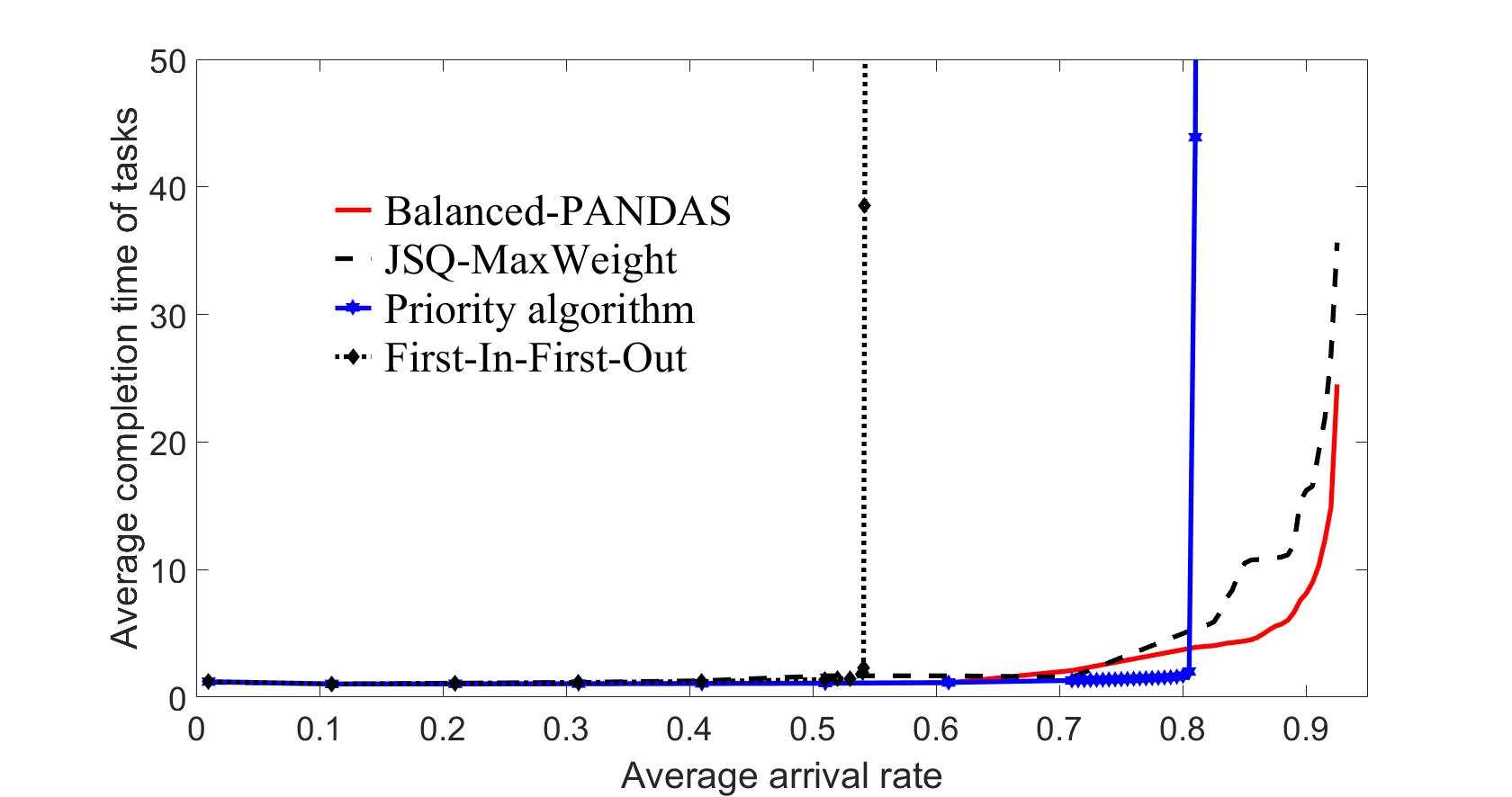}}
\subfloat[Parameters are off for 30\% higher]{\includegraphics[width = 2.495in]{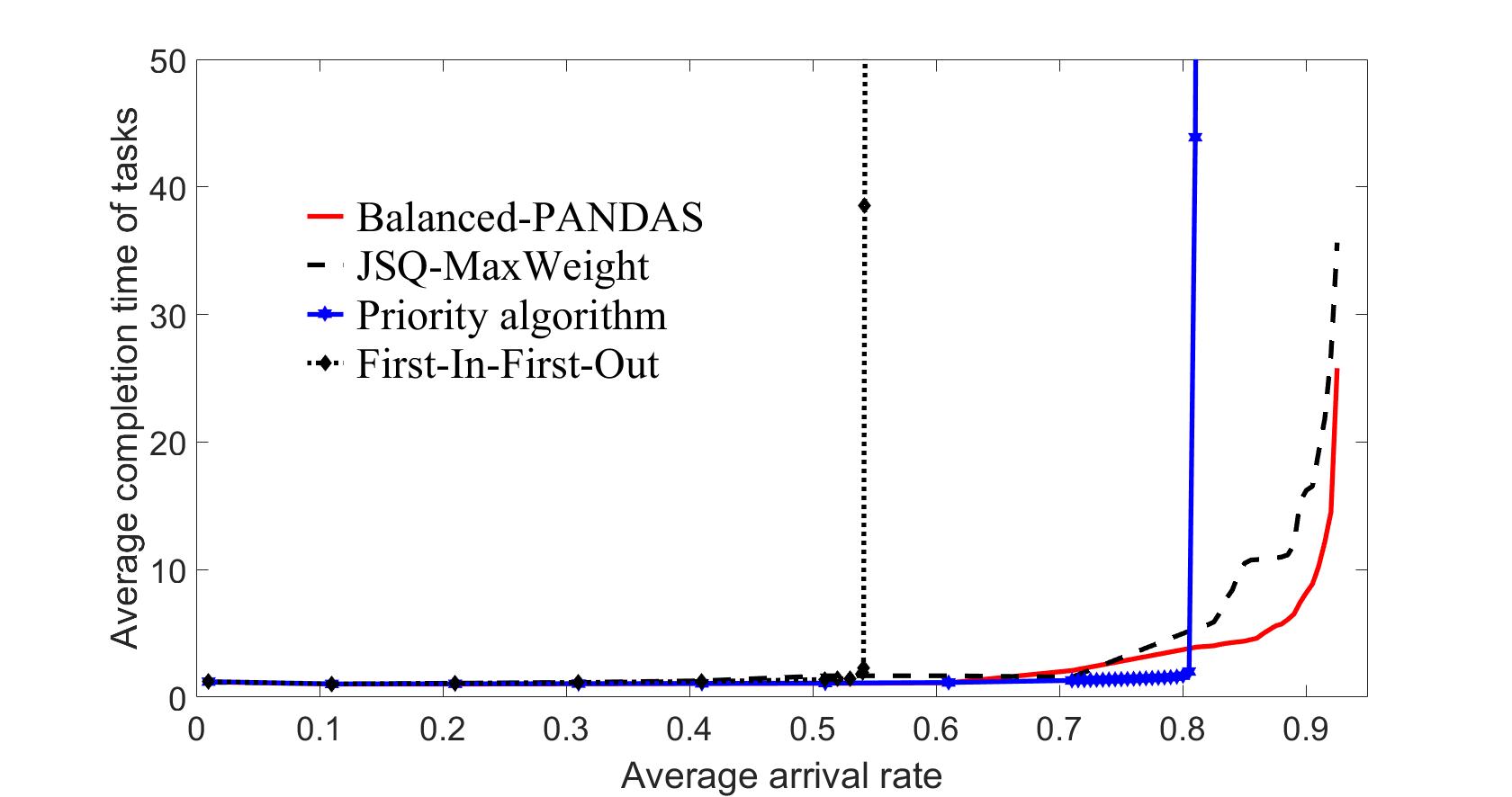}} 
\caption{Robustness comparison of algorithms when parameters are off and higher than their real values.}
\label{fig:5}
\end{figure}

\begin{figure*}
\center
  \includegraphics[width=0.75\textwidth]{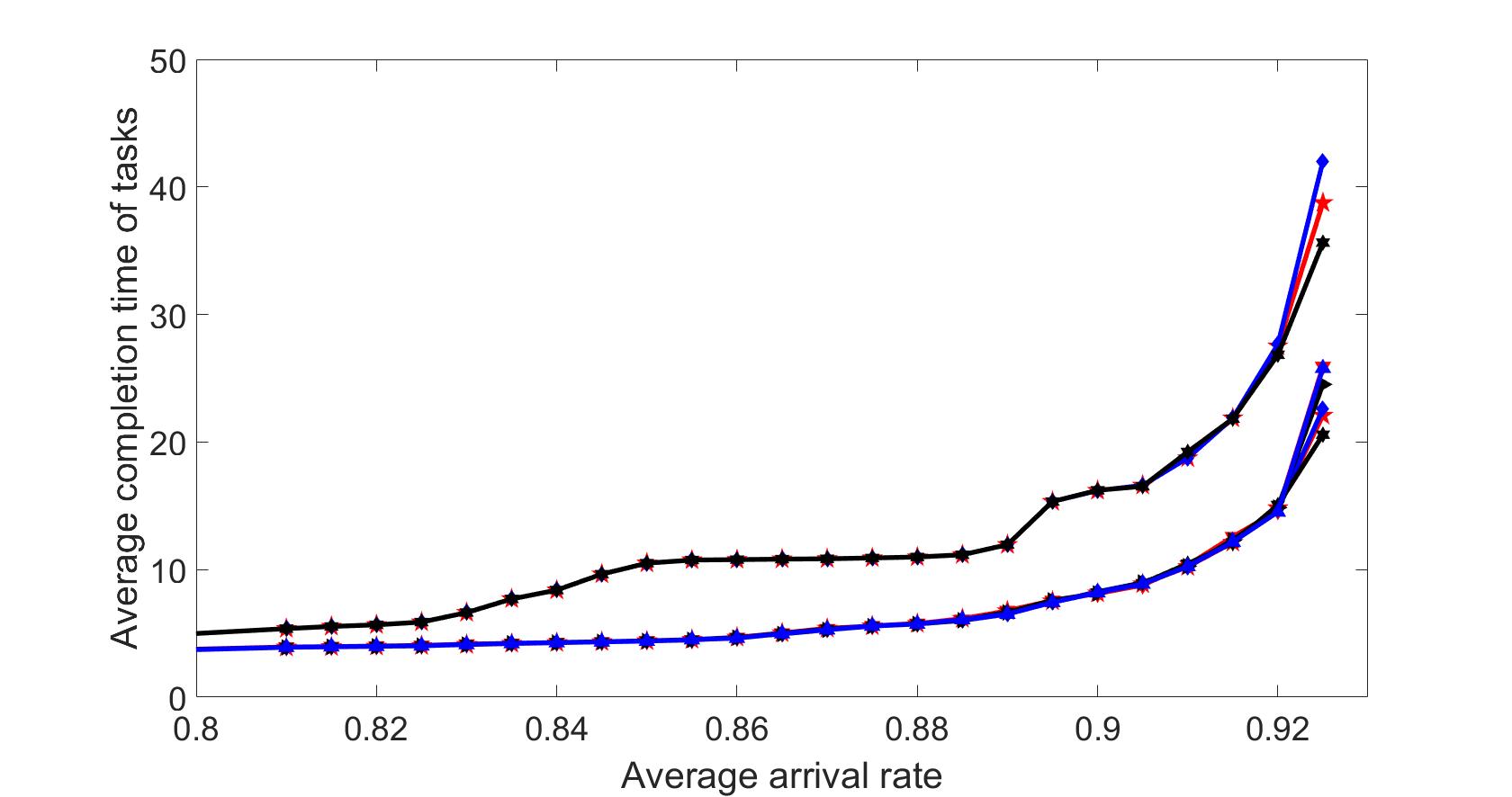}
\caption{Sensitivity comparison of Balanced-PANDAS and JSQ-MaxWeight against parameter estimation error.}
\label{fig:6}
\end{figure*}

Comparison of the algorithms when the parameters are off for some percentages, but higher than their real values are given in figure \ref{fig:5}. It is again observed that the Balanced-PANDAS algorithm has consistent better performance than the JSQ-MaxWeight algorithm. The sensitivity comparison of the Balanced-PANDAS and JSQ-MaxWeight algorithms in this case is presented in figure \ref{fig:6}.

\section{Conclusion and Future Work}
In this work, we did a literature review on both classical and state-of-the-art scheduling algorithms for the affinity scheduling problem. Data center load balancing is a special case of the affinity scheduling problem. Considering the rack structure of data centers, there are three levels of data locality. The priority algorithm that is heavy-traffic delay optimal is not even throughput optimal for three levels of data locality. The Balanced-PANDAS algorithm is the state-of-the-art in heavy-traffic delay optimality. We investigated the robustness of Balanced-PANDAS and JSQ-MaxWeight algorithms with respect to errors in parameter estimation. We observe that Balanced-PANDAS keeps its better performance even in the absence of precise parameter values versus JSQ-MaxWeight. Note that the JSQ-MaxWeight algorithm is also robust under parameter estimation errors, but it is more sensitive than Balanced-PANDAS, specially at high loads close to the boundary of the capacity region. For future work, one can use machine learning tools to estimate the system parameters and make them more precise in the meanwhile that the load balancing algorithm is working with the estimated parameters.
The scheduling algorithms presented in this work can also be applied to a vast number of applications including but not limited to  healthcare and super market models \citep{winkler1987consumerism, clower1975coordination, eisenhauer2001poor, hosseini2017mobile}, web search engines \citep{schwartz1998web, salehi2018use, broder2002taxonomy, krishna2018reducing, xie2018people}, electric vehicle charging \citep{alinia2018competitive, gan2013optimal, wang2005design, deilami2011real, chehardeh2018systematic, almalki2015capacitor, chehardeh2016remote, saadatmand2020autonomous, saadatmand2020voltage} and so on.


\newpage

\bibliography{reference}

\end{document}